# TITLE PAGE

"Self-Assembled Ge QDs Formed by High Temperature Annealing on GaAs and $Al_xGa_{1-x}As$ (001)"

William A. O'Brien,[1] Meng Qi,[1] Lifan Yan,[2] Chad A. Stephenson,[1] Vladimir Protasenko,[1] Huili (Grace) Xing,[1] Joanna M. Millunchick[2] and Mark A. Wistey[1,3]

[1] Department of Electrical Engineering, University of Notre Dame, Notre Dame, Indiana 46556

[2] Department of Materials Science and Engineering, University of Michigan, Ann Arbor, Michigan 48109-2136

[3] Corresponding author – email: mwistey@nd.edu

## ABSTRACT

This work studies the spontaneous self-assembly of Ge QDs on AlAs, GaAs, and AlGaAs by high temperature in-situ annealing in molecular beam epitaxy (MBE). The morphology of Ge dots formed on AlAs are observed by atom probe tomography, which revealed nearly spherical QDs with diameters approaching 10 nm and confirmed the complete absence of a wetting layer. Reflection high-energy electron diffraction (RHEED) and atomic force microscopy (AFM) of Ge annealed under similar conditions on GaAs and $Al_{0.3}Ga_{0.7}As$ surfaces reveal the gradual suppression of QD formation with decreasing Al-content of the buffer. To investigate the prospects of using encapsulated Ge dots for upconverting photovoltaics, in which photocurrent can still be generated from photons with energy less than the host bandgap, Ge QDs are embedded into the active region of III-V PIN diodes by MBE. It is observed that orders of magnitude higher short-circuit current is obtained at photon energies below the GaAs bandgap compared with a reference PIN diode without Ge QDs. These results demonstrate the promise of Ge QDs for upconverting solar cells and the realization of device-quality integration of group IV and III-V semiconductors.

**INTRODUCTION**

Alternating growth between lattice-matched Ge and (Al)GaAs [1-4] has been extensively investigated, and has long been pursued as a bridge between group IV semiconductors and III-V zincblende materials. Growth of single IV to III-V transitions can be fairly robust with vicinal substrates and thick buffer layers, as used in several multijunction solar cells [5-8]. However, some issues remain challenging, particularly mitigating the formation of anti-phase domains when switching from the diamond to the zincblende lattice without a buffer layer, or on exact-cut (100) substrates.

Lateral epitaxial overgrowth is one approach to reduce antiphase domains, and smaller features such as self-assembled quantum dots (SAQDs) would be easier to overgrow. But most QDs, including Ge on Si, form by the Stranski-Krastanov (S-K) growth mode, which leaves a continuous wetting layer. The wetting layer breaks the continuity of the zincblende lattice, so subsequent III-V growth above the QDs again suffers from anti-phase domains. Even so, Ge QDs are interesting for their optoelectronic properties, such as light emission orders of magnitude higher than from equivalent quantum wells on Si. And the long excited state lifetimes in Ge ($\tau_n \sim$ 1 ms) offer compelling device opportunities that are not otherwise available in conventional III-V semiconductors, such as sequential upconversion. Also, QD systems formed by the S-K method generally require strain, which can limit the device thickness before strain relaxation and defects form.

We have previously demonstrated an alternative process for realizing Ge QDs on AlAs by high temperature annealing without a wetting layer or strain. When the dots are overgrown by AlAs, this method also enables III-V's and group IV's to be grown together seamlessly, without forming anti-phase domains or visible defects in cross-sectional TEM.

In this work, we observe striking differences in Ge QD formation depending on the substrate surface using atom probe tomography (APT), atomic force microscopy (AFM), and reflection high energy electron diffraction (RHEED). In addition, photocurrent in PIN structures shows significant absorption by the QDs.

## MATERIALS AND METHODS

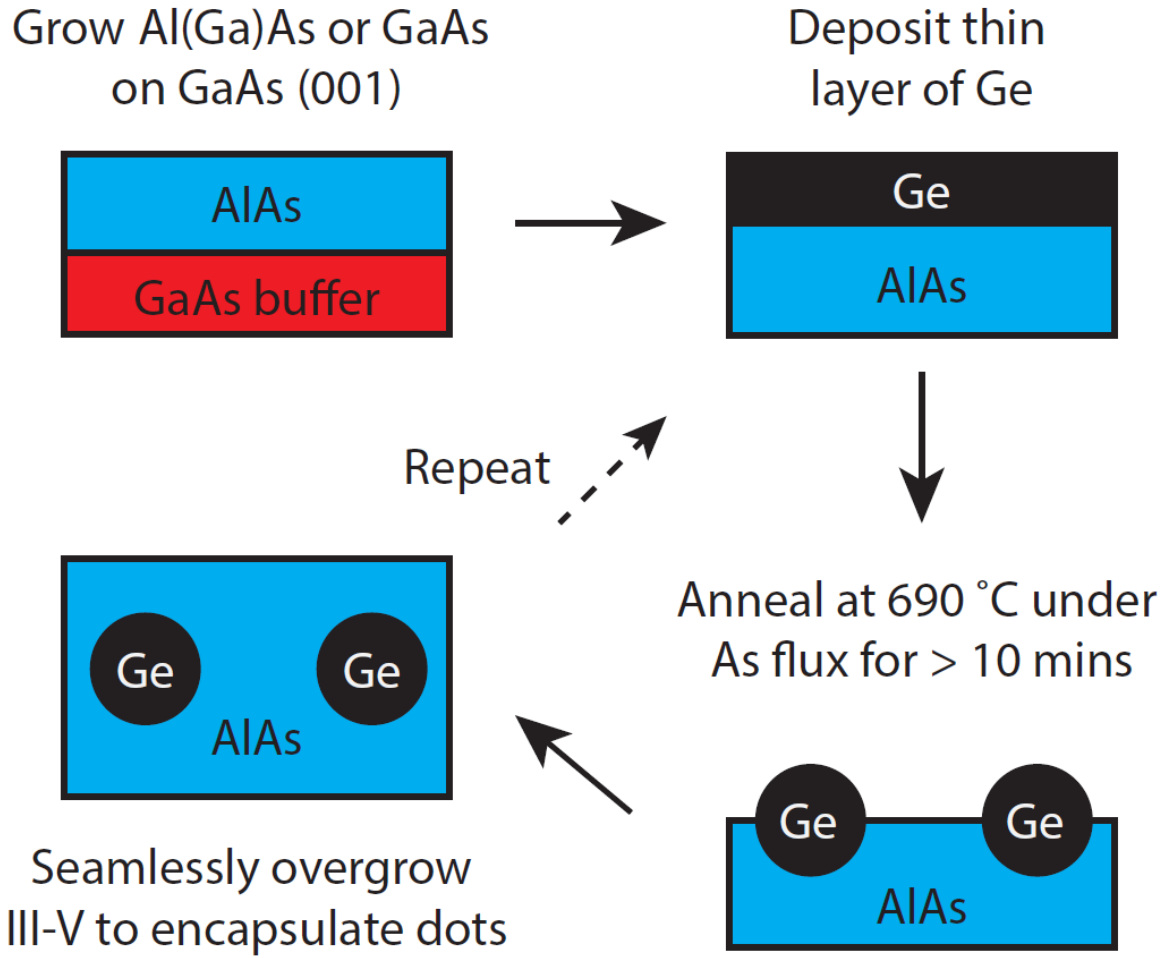

Fig. 1. Process flow for Ge QD formation on AlAs. After dot encapsulation, the process can be repeated indefinitely.

The process for Ge QD formation, depicted in Fig. 1, was as follows. The native oxide on GaAs was desorbed at 630-650 °C for at least 10 minutes, followed by a GaAs buffer. For some samples, a smoothing superlattice of AlAs/GaAs was grown. Next a layer of AlAs, AlGaAs, or GaAs was grown on which the Ge would be deposited. A thin layer of Ge on the order of a unit cell (0.56 nm) was evaporated from a solid effusion source, either at 410 °C (thermocouple) without As flux or at 690 °C with As flux to prevent AlAs decomposition. For all samples, the substrate temperature was then held at 690 °C under $As_2$ flux to anneal the Ge film for 10 minutes. For samples with multiple layers of Ge, including the PIN samples (discussed below), the AlAs and Ge depositions and anneal were repeated for each layer. When grown on AlAs, the outcome of this process was the spontaneous formation of roughly spherical Ge dots. Longer annealing times will ripen the QDs into larger dots while maintaining a spherical symmetry, even for dots exceeding 10 nm in diameter. The density of the dots can be controlled by the initial thickness of the Ge film. Once the QDs are formed, they are capped by AlAs. This process can then be repeated for additional periods of QD layers. The final structure is typically terminated by a GaAs cap to prevent oxidation of the AlAs.

The APT specimens were prepared using the site-specific lift-out technique in FEI Helios Nanolab 650 Dualbeam FIB [9]. The specimen were milled into tips with diameter less than 100 nm. The APT analysis was done in a Cameca LEAP 4000X at temperature of 25K with assistance of laser power of 0.12 pJ. The laser power was tuned to reach the best Al/As atomic ratio to ensure the most accurate determination of the elemental compositions.

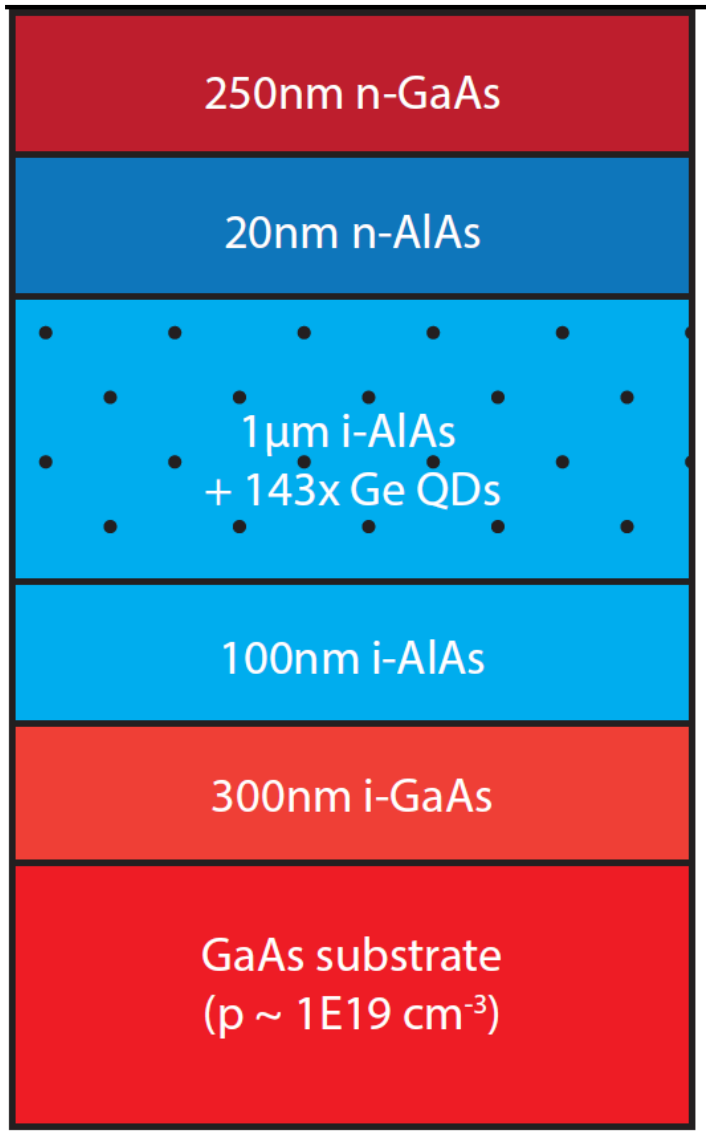

Fig. 2. Schematic of epitaxial structure for Ge QD PIN diode.

PIN diodes were fabricated from the epitaxial structure shown in Fig. 2. An unintentionally doped (NID) buffer of 300 nm GaAs and 100 nm AlAs was grown on a *p+* GaAs substrate (Zn-doped, p ~ 1x10$^{19}$ cm$^{-3}$). Ge was then deposited and annealed at 690 °C under an As flux 6.7x10$^{-6}$ torr for 10 minutes to form QDs. With 7 nm NID AlAs spacers grown between QD layers, a total of 143 repetitions are performed. This resulted in a 1 µm layer of AlAs with embedded Ge dots. A control PIN structure was also grown without Ge dots in the i-AlAs region, but with the same As annealing steps in case background impurities are incorporated during the 10 minute high temperature anneal. Finally, an n-AlAs and n-GaAs cap were deposited to complete the PIN structure.

For diode fabrication, first an n-ohmic metal stack composed of 40 nm Au, 12 nm Ge, 12 nm Ni, and 200 nm Au was evaporated through a photoresist (PR) mask, followed by liftoff and a one minute rapid thermal anneal in $N_2$ at 400 °C. Then PR patterning was used to define mesa structures to control the effective diode area, followed by a wet etch in 1:1:10 $H_3PO_4/H_2O_2/H_2O$, which penetrated to the substrate. After the PR was removed, ohmic contacts to the p-GaAs substrate, composed of 40 nm Ti, 40 nm Pt, and 200 nm Au were evaporated on the backside of the wafer. Fig. 3(a) and 3(b) shows the simulated band diagram for the Ge QD and control PIN structure, assuming negligible traps and interdiffusion of AlAs and Ge in the intrinsic region. In the case of the Ge QD structure, the band diagram represents a cross section through a vertical stack of Ge QDs separated by 7 nm AlAs layers. The inset of Fig. 3(a) shows the conduction band near the first three periods of QDs in the structure, which are shown to be populated with electrons in equilibrium.

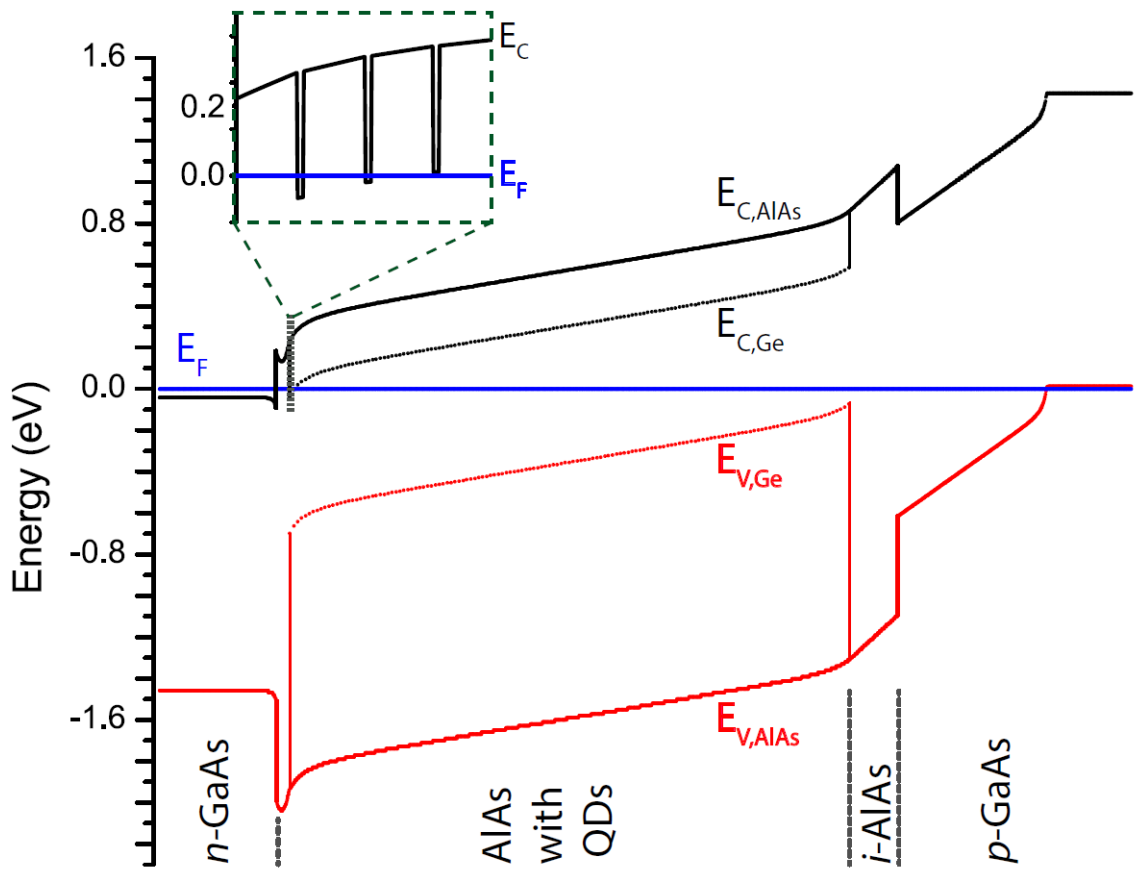

Fig. 3. Band diagram for Ge QD PIN structure. For clarity, only the tops and bottoms of the QD wells are shown, except for 3 QDs near the surface of the structure shown in the inset.

## MORPHOLOGY RESULTS AND DISCUSSION

Fig. 4 shows the Atom Probe Tomography (APT) of the structure annealed on AlAs and shows elliptical QDs with an average size of 400 nm$^3$. The distribution of the dots is not Gaussian but exponential, indicating that there is a very large number of small QDs and a long tail in the distribution for larger QDs, consistent with a continuous nucleation process. No wetting layers are observed. The shape of the QDs are distorted in the APT image compared to how they would appear in a Transmission Electron Microscope (TEM) image [10] due to the fact that Ge and AlAs mill at slightly different rates. QDs that are relatively large are arranged in layers, consistent with the growth sequence, and small QDs are visible between the layers.

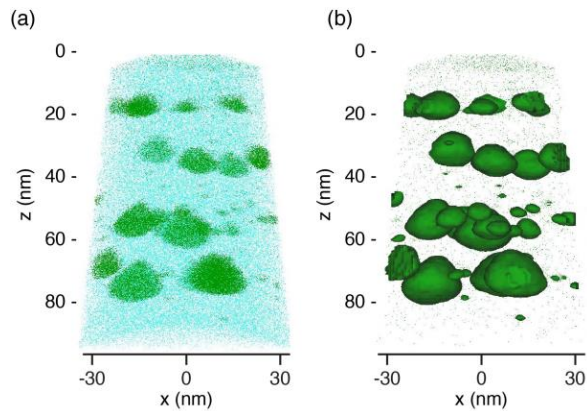

Fig. 4. Atom probe tomography. (a) Concentrations of Ge (green) and As (aqua) in a stack of 4 layers of Ge SAQDs separated by AlAs. (b) Ge 3D profile of same region generated by proxigram analysis [11] with 6% concentration threshold. Note lack of wetting layer.

At the end of the anneal, as shown in TEM [10] and the APT images in Fig. 4, it would appear the Ge dots have sunken into the AlAs matrix. Furthermore, Ge annealing has been investigated on GaAs and Al$_{0.3}$Ga$_{0.7}$As surfaces. In contrast to growth on AlAs[10], in which the QDs were nearly spherical, Ge nanostructures on AlGaAs in Fig. 5(c) appear to be better described as flat disks, with a slight depression at the center. These are similar to structures formed by droplet epitaxy [12], providing evidence for liquid droplet formation. Furthermore, there is no indication of Ge nanostructure formation on GaAs as shown in Fig. 5(a). RHEED images taken before and after annealing on GaAs are shown in Fig. 6. Though the streaks are somewhat less continuous, it indicates the surface after annealing is still relatively smooth, corroborating AFM images which indicate Ge QD formation is suppressed.

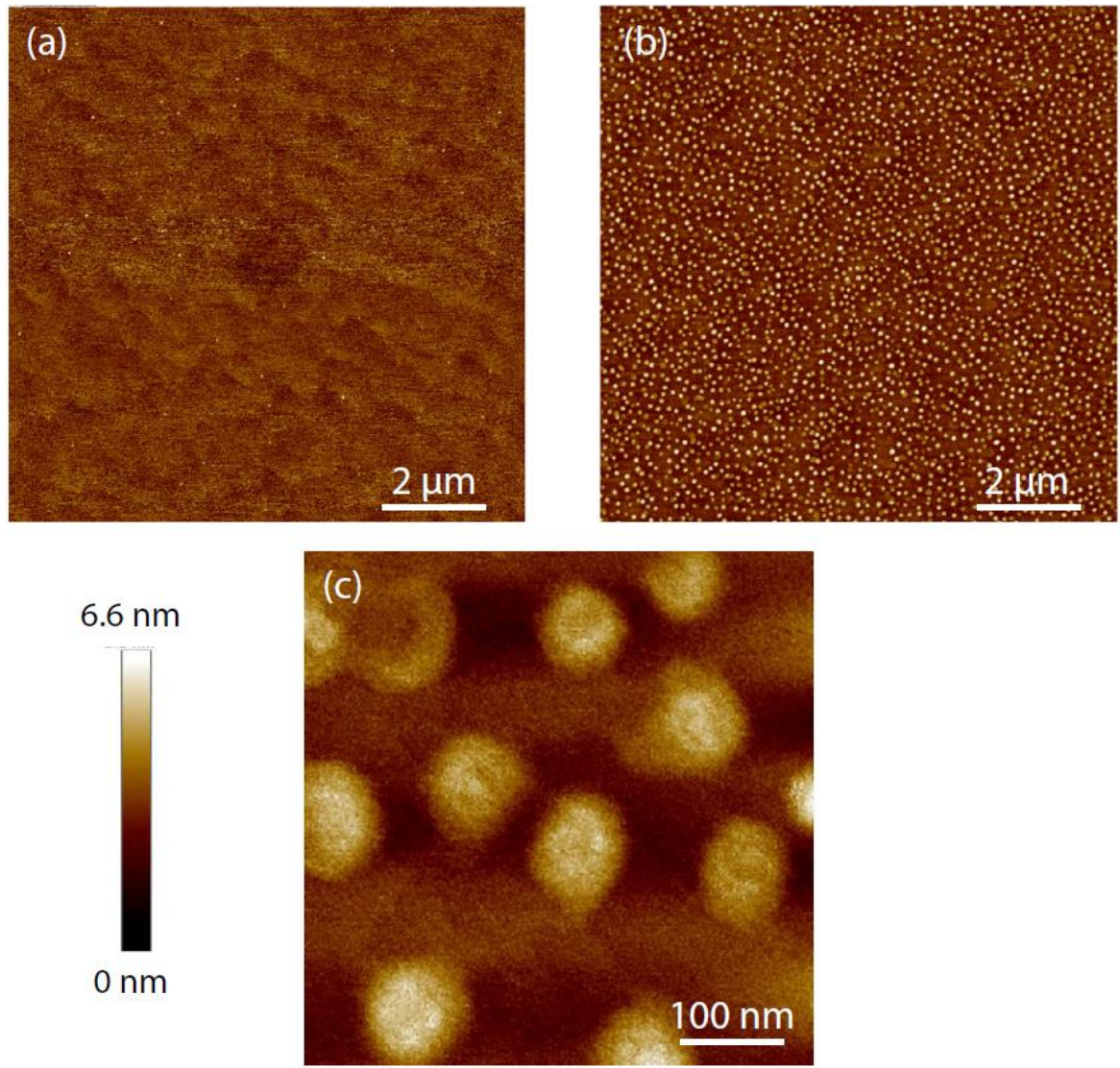

Fig. 5. Atomic force microscopy (AFM) after high temperature annealing of Ge on (a) GaAs and (b) & (c) on $Al_{0.3}Ga_{0.7}As$.

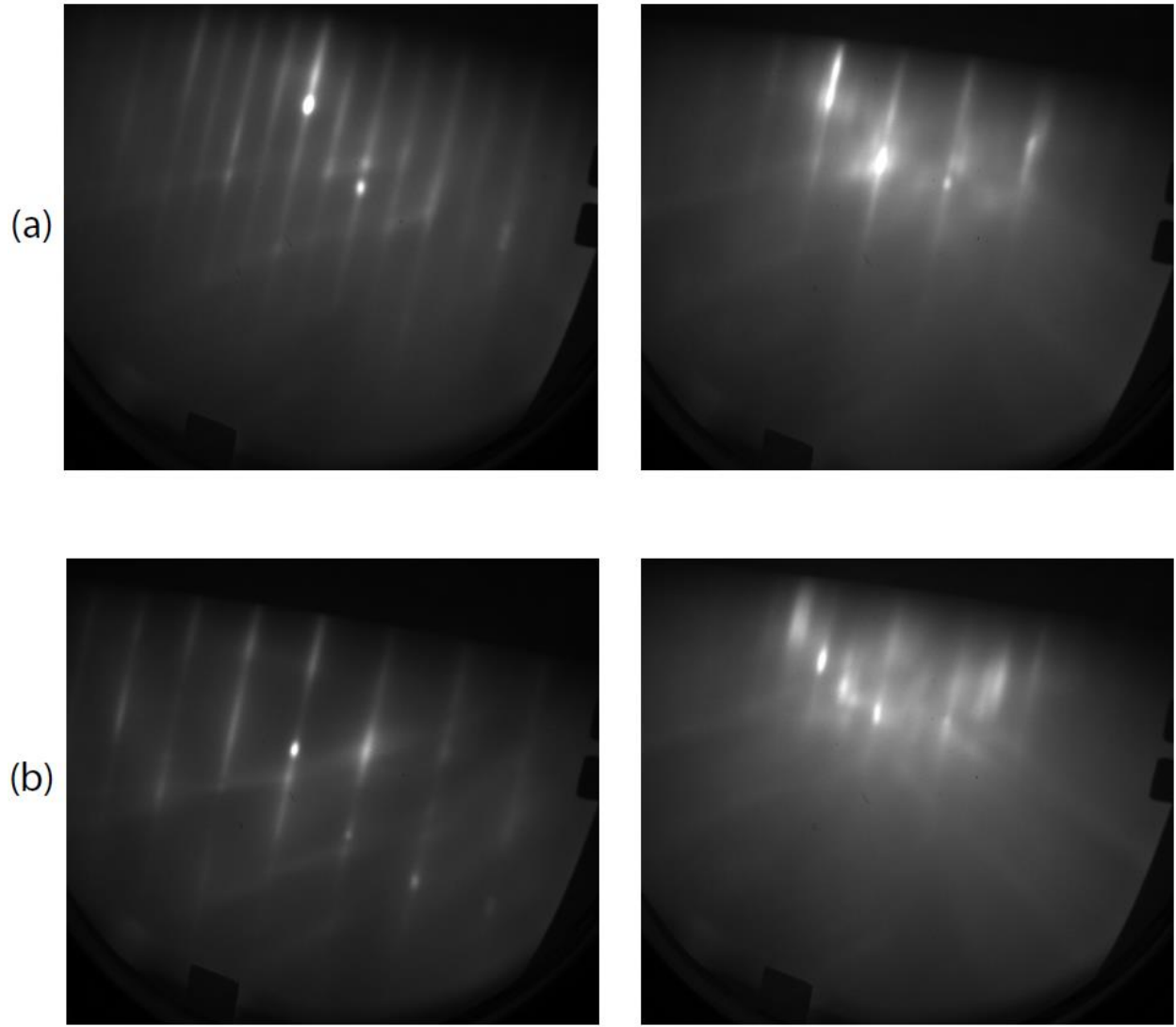

Fig. 6. RHEED pattern at 750 °C (thermocouple), (a) before and (b) after high temperature annealing of Ge on GaAs. The transition from a 2x3 to 2x5 reconstruction is observed after the process.

**Possible origins of Ge QD formation**

Unlike QDs formed by the S-K method, the formation of Ge QDs on lattice-matched AlAs is not driven by strain energy. The surface free energy clearly plays a significant role, since the Ge formed nanostructures on AlGaAs but did not form any structures on GaAs under the same conditions. The observation of Ostwald ripening with annealing time strongly indicates the formation mechanism is driven by surface energy as well. The formation of spherical dots that are partially sunken into the underlying layer may be explained if the Ge formed a eutectic liquid with AlAs upon annealing, followed by the precipitation of a Ge-AlAs alloy dot upon cooling.

A phase diagram for the AlAs-Ge system [13] indicates a eutectic point at 735 °C, with a Ge composition of 57%. This suggests that it is possible for spherical dots to be formed by melting of the layer. However, the maximum annealing temperature was only 690 °C, calibrated by band edge thermometry. This temperature is below the bulk eutectic temperature. On the other hand, much higher surface energies due to large surface to volume ratios in nanostructures can significantly lower the eutectic temperature and composition [14]. This is based on considering the Gibbs free energy for each phase, in which surface energy can significantly shift the onset of phase transitions from equilibrium points established in the bulk phase diagram. Upon cooling, the core of the QDs would contain Al and As near the solubility limit in Ge, which is about 2% for bulk alloys at 690 °C.

However, analysis of atom probe data in Fig. 7 indicates that the cores of the QDs are at least 80% Ge and less than 14% Al, which is inconsistent with the eutectic composition. An explanation for this may be that the solubility of the AlAs is enhanced for nanoparticles. The boundaries of the QD in the proxigram analysis appear to be graded over a distance of 2-3 nm, as might be expected. However, the milling rates for the Ge and AlAs were somewhat different during acquisition of the APT data, which would result in a distortion of the apparent nanoparticle shape and interfacial abruptness.

With these considerations in mind, a possible formation mechanism can be envisioned as follows. The radius of curvature required for the liquid phase to become stable at lower temperatures can be catalyzed by asperities from atomic-scale roughness present in deposited Ge thin films. Upon reaching 690 °C, the effective eutectic temperature for small nuclei is reached, which spontaneously melts the layer. The liquid layer forms droplets to reduce the surface energy, and because AlAs is somewhat soluble in Ge, dissolves some of the underlying layer in a manner similar to droplet epitaxy of GaAs QDs in AlGaAs. The dots would then crystallize upon cooling.

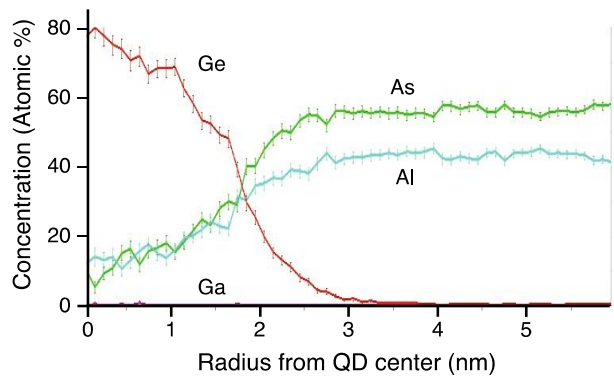

Fig. 7. Proxigram analysis [11] of atom probe tomography data.

## PHOTOCURRENT RESULTS AND DISCUSSION

To study the potential for sequential absorption from multiple photons in the QD structures, we examine the short-circuit photocurrent, after removing the optical system response, near the GaAs absorption edge in Fig 8. The photocurrent from photon energies just above the GaAs bandgap originates from absorption in either the $n^+$ GaAs cap, the Ge QDs in the $i$-AlAs region, or the GaAs below the $i$-AlAs region. We estimate that less than 25% [15] of this light is absorbed in the cap.

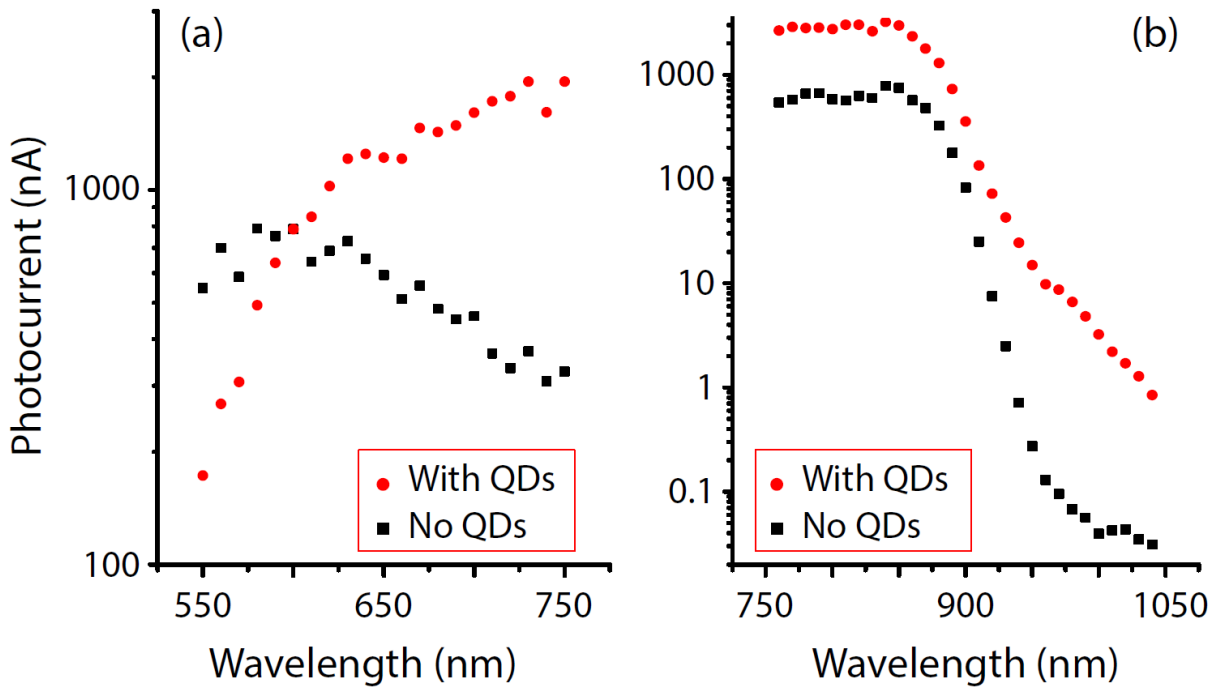

Fig. 8. Photoconductance measurements of Ge QDs at zero applied bias (a) near the AlAs absorption edge and (b) near the GaAs absorption edge.

For electrons and holes drifting through the *i*-AlAs/QD region, carriers that fall into the dots can be re-excited out of the QDs by absorbing another photon. One benefit of using an AlAs matrix, with an indirect band gap, is long diffusion lengths attainable and the ability to collect current without significant band-to-band recombination.

At photon energies below the GaAs bandgap, electron-hole pair (EHP) generation is expected to occur almost exclusively within the Ge dots. Subsequent photons can then be absorbed by excited carriers in the QDs in a bound-to-continuum (B-C) transition, enabling them to be collected as external current. In addition, the B-C transition in the conduction band is also favorable for Ge QDs near the surface, in which the dots are already populated with electrons, according to the equilibrium band diagram.

When compared to a control PIN diode, it is clear that orders of magnitude more photocurrent is generated by the Ge QD PIN below the GaAs bandgap. This can be attributed to EHP generation in the Ge dots, aided by the B-C transitions. When looking above the GaAs absorption edge in Fig. 8(b), near the onset of absorption in the AlAs, it is observed that the control PIN overtakes the Ge QD diode in photocurrent. This can be attributed to prevalence of carrier trapping in the dots and a reduction in QE from photons absorbed in Ge.

It should be noted that the control sample was grown at a later date and shows indications of somewhat lower material quality. In particular, photocurrent is generally lower, including photon energies just above both the GaAs and AlAs bandgaps where photocurrent should rise sharply regardless of Ge. We believe the decreased internal quantum efficiency (IQE) indicates a slightly higher trap density and increased nonradiative recombination in the control, but not sufficiently high to change the prevailing optical transitions. This is borne out by a lower overall photocurrent without additional features in the spectrum. However, the increased recombination in the control sample precludes quantitative extraction of IQE in the Ge QDs.

## CONCLUSIONS

Ge QDs integrated into a PIN diode have shown a nearly two orders of magnitude increase in photocurrent below the bandgap of the III-V constituents, over an identical PIN structure without dots. This suggests Ge QDs embedded in GaAs-based photovoltaics may be useful for capturing wasted infrared light through a two-step absorption process. This produces extra current from sub-bandgap photons which can be collected at the higher voltage of the matrix material. Future work will determine to what extent the open circuit voltage is affected by the presence of the Ge dots in the device active region.

Atom probe and AFM measurements show nearly spherical Ge QDs self-assemble on AlAs after anneal under As. Ge QDs did not form on GaAs surfaces. The Ge formed nanorings rather than spherical dots on AlGaAs, which supports the liquid droplet epitaxy model. Future experiments are directed toward providing a detailed model for the mechanism of formation, as well as the composition of the QDs and bonding at the Ge/(Al)GaAs interfaces.

## ACKNOWLEDGEMENTS

This work was supported by the National Science Foundation under CBET-1438608, and by the Notre Dame Integrated Imaging Facility and MIND center. The authors declare no financial conflicts of interest. The authors also thank Alexander Mintairov for technical assistance.## REFERENCES

1. P. M. Petroff, A. C. Gossard, A. Savage and W. Wiegmann, *J. Cryst. Growth*, 46, 2, pp. 172-178 (1979)
2. T. S. Kuan and C.-A. Chang, *J. Appl. Phys.*, 54, 4408 (1983)
3. M. K. Hudait, Y. Zhu, N. Jain and J. L. Hunter, *J. Vacuum Sci. Technol. B*, 31, 1 (2013)
4. R. Fischer, W. T. Masselink, J. Klem, T. Henderson, T. C. McGlinn, M. V. Klein, H. Morkoc, J. H. Mazur and J. Washburn, *J. Appl. Phys*, 58, 374 (1985)
5. M. A. Stan, P. R. Sharps, N. S. Fatemi, F. Spadafora, D. Aiken and H. Q. Hou. "Design and production of extremely radiation-hard 26% InGaP/GaAs/Ge triple-junction solar cells," Conference Record of the 28$^{th}$ IEEE PVSC, Sept. 15-22, Anchorage, AK, pp. 1374-1377 (2000)
6. T. Takamoto, M. Kaneiwa, M. Imaizumi and M. Yamaguchi, *Prog. Photovoltaics*, 13, 495 (2005)
7. M. J. Archer, D. C. Law, S. Mesropian, M. Haddad, C. M. Fetzer, A. C. Ackerman, C. Ladous, R. R. King and H. A. Atwater, *Appl. Phys. Lett.*, 92, 103503 (2008)
8. R. R. King, D. C. Law, K. M. Edmondson, C. M. Fetzer, G. S. Kinsey, H. Yoon, R. A. Sherif and N. H. Karam, *Appl. Phys. Lett.*, 90, 183516 (2007)
9. K. Thompson, D. Lawrence, D. J. Larson, J. D. Olson, T. F. Kelly and B. Gorman, *Ultramicroscopy*, 107, 2, 131-139 (2007)
10. M. Qi, C. A. Stephenson, V. Protasenko, W. A. O'Brien, A. Mintairov, H. Xing and M. A. Wistey, *Appl. Phys. Lett.*, 104, 073113 (2014)
11. O. C. Hellman, J. A. Vandenbroucke, J. Rusing, D. Isheim and D. N. Seidman, *Microsc. Microanal.*, 6, 437-444 (2000)
12. K. Reyes, P. Smereka, D. Nothern, J. M. Millunchick, S. Bietti, C. Somaschini, S. Sanguinetti and C. Frigeri, *Phys. Rev. B*, 87, 165406 (2013)
13. F. R. Schmid, Ternary Alloys (VCH, 1994), Vol. 9, p. 97
14. H. Adhikari, A. F. Marshall, I. A. Goldthorpe, C. E. D. Chidsey and P. C. McIntyre, *ACS Nano*, 1, 5, 415-422 (2007)
15. H. C. Casey Jr., D. D. Sell and K. W. Wecht, *J. Appl. Phys.*, 46, 250 (1975)
Page 14